# Deep Attentive Generative Adversarial Network for Photo-Realistic Image De-Quantization

Yang Zhang, Changhui Hu, and Xiaobo Lu

***Abstract*—Most of current display devices are with eight or higher bit-depth. However, the quality of most multimedia tools cannot achieve this bit-depth standard for the generating images. De-quantization can improve the visual quality of low bit-depth image to display on high bit-depth screen. This paper proposes DAGAN algorithm to perform super-resolution on image intensity resolution, which is orthogonal to the spatial resolution, realizing photo-realistic de-quantization via an end-to-end learning pattern. Until now, this is the first attempt to apply Generative Adversarial Network (GAN) framework for image de-quantization. Specifically, we propose the Dense Residual Self-attention (DenseResAtt) module, which is consisted of dense residual blocks armed with self-attention mechanism, to pay more attention on high-frequency information. Moreover, the series connection of sequential DenseResAtt modules forms deep attentive network with superior discriminative learning ability in image de-quantization, modeling representative feature maps to recover as much useful information as possible. In addition, due to the adversarial learning framework can reliably produce high quality natural images, the specified content loss as well as the adversarial loss are back-propagated to optimize the training of model. Above all, DAGAN is able to generate the photo-realistic high bit-depth image without banding artifacts. Experiment results on several public benchmarks prove that the DAGAN algorithm possesses ability to achieve excellent visual effect and satisfied quantitative performance.**

## I. Introduction

Bit-depth (BD) [1] represents the number of bits to store intensity values in image RGB channels, resulting in the effects on the image dynamic range and display quality. With the rapid development of the display technology, the BD of most display devices is 8 or higher bit. However, many multimedia tools still cannot achieve this standard, equipped with even lower BDs. Thus, image de-quantization or bit-depth bit-depth (HBD) image recovering from the corresponding low bit-depth (LBD) image, filling the least significant bits (LSBs) [12] between HBD and corresponding LBD image.

This work is done when the first author visits the University of Technology, Sydney. Y. Zhang is with the School of Automation, Southeast University, Nanjing 210096, China; the Faculty of Engineering and Information Technology, University of Technology Sydney, Ultimo, NSW 2007, Australia; the Key Laboratory of Measurement and Control of Complex Systems of Engineering, Ministry of Education, Nanjing 210096, China (e-mail: zhangyang201703@126.com).

CH.Hu and X. Lu are with the School of Automation, Southeast University, Nanjing 210096, China; Key Laboratory of Measurement and Control of Complex Systems of Engineering, Ministry of Education, Nanjing 210096, China (email: xblu2013@126.com).

Until now, many methods focus on image de-quantization have been put forward. The traditional image de-quantization algorithms, including zero-padding method (ZP) [4], bit replication method (BR) or ideal gain method (MIG) [5], directly realize pixel-based mapping between LBD and HBD image. However, due to their context independent property and regardless of spatial correlation, unnatural artifacts which affect the visual quality are the common products.

Aiming at remedying the shortcomings of the traditional methods, the following proposed de-quantization methods, such as flooding based method (FBM) [6], minimum risk based classification (MRC) [7], graph signal processing (GSP) [8], maximum-a-posteriori (MAP) [9], IPAD [10], etc, mainly concentrate on the banding artifact removal. However, although they have handled the unnatural artifacts and banding effects to some extent, the shortcomings such as sensitive to isolated noise, regardless of useful high-frequency information and distorted recovering of missing LSBs remain to be addressed.

Later, some CNN based methods are proposed [11-15]. Yang et.al [12] proposed the BDEN model, which combines traditional debanding strategy into trained residual blocks based network. The corresponding subjective quality is improved compared to traditional methods. In [13], ICH-CNN is proposed to utilize the original U-net structure to realize image de-quantization task, aiming at deleting banding artifact. Admittedly, deep learning based approach is conductive to modeling high-frequency features. Yet, for the unnatural artifacts and banding effects eliminating ability, they always cannot achieve the excellent performance of tradition method.

Recently, the community has witnessed rapid developments of generative models for image data based on the adversarial learning framework. The GAN [16] models can reliably produce high quality images. This indicates that a trained GAN model implicitly contains information about the characteristics of the natural image distribution.

It is therefore natural to postulate that a trained GAN model might be helpful in recovering useful information from LBD image based on the modeled natural image distribution. It is especially true when the available observation contains the content of the image-information of the LBD image and the task is to recover the adaptive content of the image-information to construct photo-realistic HBD images.

Inspired from previous super resolution researches [23], [24] which focus on super-resolution (SR) on spatial resolution. In this work, we consider to utilize GAN framework to realize SR on image intensity resolution, which is orthogonal to the spatial resolution, realizing image de-quantization with photo-realistic texture recovering. Fig. 1 illustrates the significant parameters, including spatial resolution and intensity resolution, which affect the image quality from the perspective of information recording of sensors in multimedia tools. Our intuition is that, GAN based network can be trained to generate certain HBD image pattern by presenting the network with a set of well-structured LBD-HBD image pairs as well as rationally distributed content loss.

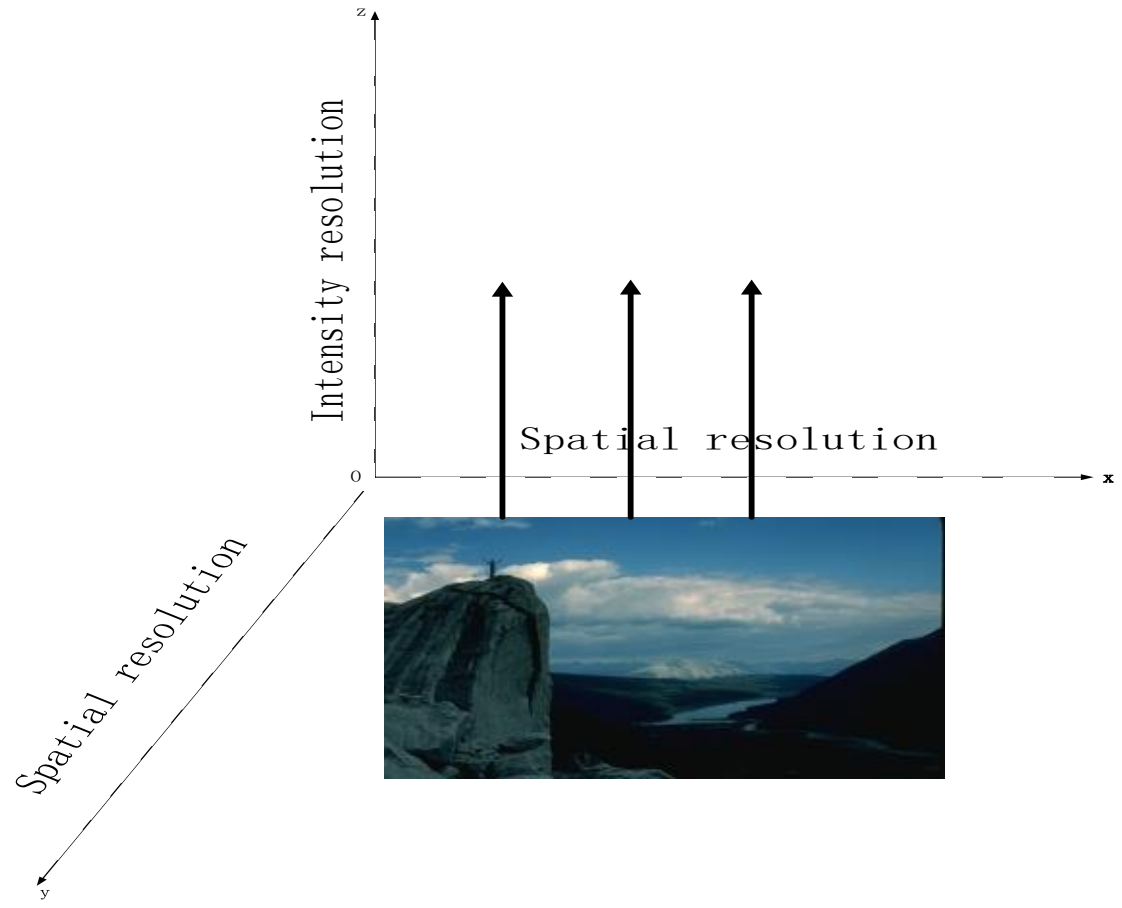

Fig. 1. The significant parameters of an image from the perspective of information recording.

Compared to other state-of-the-art de-quantization methods focused on color gradation smoothing, noise eliminating and banding artifacts reducing. DAGAN puts photo-realistic high-frequency details recovering into effect, additionally. The example photo-realistic HBD image generated by DAGAN model is shown in Fig. 2.

Overall, our contributions are fourfold:

1) We first attempt to utilize GAN based model into image de-quantization, realizing SR on intensity resolution, which is orthogonal to the spatial resolution, generating the photo-realistic HBD image via an end-to-end learning pattern. DAGAN model can well handle the banding artifacts without relying on any traditional image processing method.
2) We propose the Dense Residual Self-attention (DenseResAtt) module, which enables the network to pay attention to useful information. Then, the sequential DenseResAtt modules as well as alternating long and short skip connections work together to form deep attentive network with superior discriminative learning ability, recovering high-frequency information and bypassing low-frequency information.
3) Armed with self-attention mechanism, DAGAN network realizes close connection of fine details from the comprehensive perspective as well as complicated geometric constraint on the feature maps, learning more effective information to promote the recovery of discriminative details. In this way, good visual performance for the generating HBD image is achieved.
4) We define a specified content loss, which is more sensitive to human perception, together with the adversarial loss to encourage the generating HBD result perceptually hard to be distinguished from the ground truth HBD image, realizing photo-realistic image de-quantization.

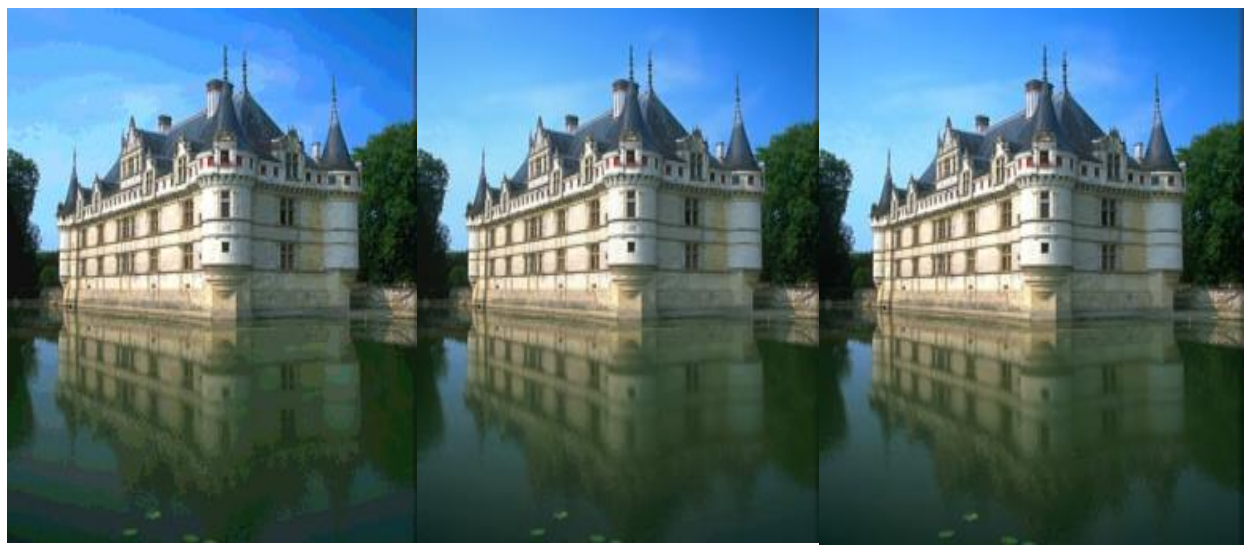

(a) LBD image (b) Result of DAGAN (c) Ground Truth

Fig. 2. Reconstructed image (b) is almost indistinguishable from ground truth image (c)

## II. Related work

### A. Design of GAN

This research is related to the GAN [16] framework, which trains generator (G) and discriminator (D) via a min-max two player game to learn image prior. GANs have achieved great success in various image tasks, including image-to-image translation [17-20], image generation [21], [22], [65-71], image super-resolution [23, 24], text-to-image synthesis [25-27], image editing [28], [29] and image inpainting [30], [31].

The basic GAN module is constructed by two independent components, the G network and D network. D network learns judge the generating sample is real or fake. Contradictorily, G network learns to generate fake image to fool D. In the course of this contradictory game, adversarial loss is modeled to optimize corresponding tasks. Even though the training process of GAN is sensitive to hyper-parameters. The corresponding optimization strategies, including design modified network architectures [32], [33], setting the learning objectives as well as dynamics [34-36], appending regularization techniques [37], [38] and combining heuristic methods [39], [40], are put in to research to make the GAN training dynamics stable.

In SRGAN [23], GAN model is adopted to transfer artistic style to photos, with impressive success. Given a low spatial resolution image, SRGAN utilizes a GAN model to produce a high quality image with enhanced spatial resolution.

Similar but alternative, this research employs GAN structure to align the distribution of intensity resolution among generating images with ground truth images. Aiming at generating HBD image with recovered photo-realistic texture and deleted banding artifacts.

### B. Self-attention mechanism

Recently, the attention module, which is embedded into network to outstand global dependencies [41-44] has become an efficient part of a range of techniques. Especially, self-attention mechanism [45], [46], which calculates the response at a position in the feature by referring to all positions within the same feature, performs excellent effect for the GAN model.

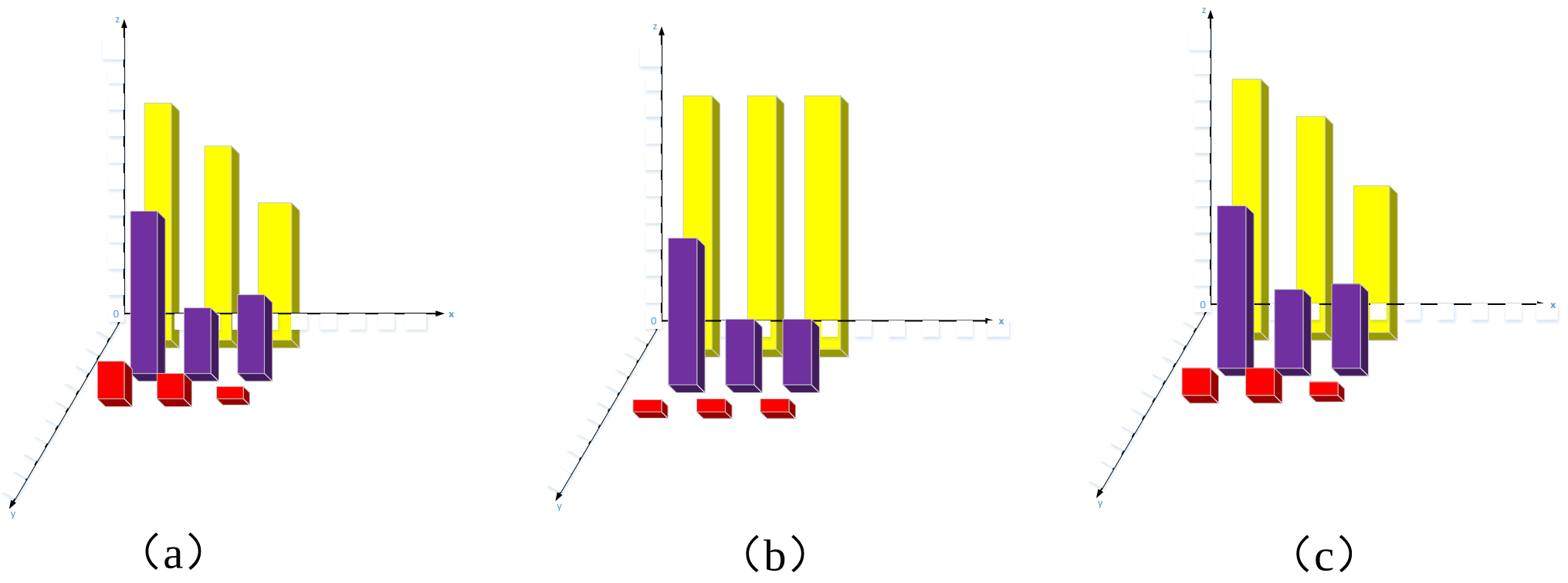

Fig. 3. The SR process of pixel intensity. (a) local pixel intensity distribution in ground truth HBD image. (b) local pixel intensity distribution in LBD image. (c) local pixel intensity distribution in reconstructed HBD image.

Until now, the self-attention mechanism has been used in image generation [47], [48] and video analyzation [49]. Parmar et al. [50] propose a self-attention based autoregressive model to automatically generate vivid image. Vaswani et al. [51] employ self-attention model into machine translation framework to get state-of-the-art performance. Zhang et al. [52] put forward the Self-Attention Generative Adversarial Networks (SAGAN) to model the internal representations of images, focusing on the image generation research. Wang et al. [49] use the self-attention module as a non-local operation to produce the spatial-temporal dependencies in video processing task. Qian et al. [53] propose a novel recurrent network to produce visual attention, challenging the raindrop eliminating task.

In spite of this progress, self-attention mechanism has not yet been explored in the image de-quantization task. Aiming at getting high quality HBD image, self-attention mechanism is embedded into the DenseResAtt module in our DAGAN network, learning more effective information to promote the recovery of discriminative details.

### C. *Image de-quantization algorithm*

This research hold the opinion that the image quality can be enhanced on spatial resolution as well as intensity resolution, respectively. Super resolution (SR) is the commonly research technique for image quality enhancement, which focuses on spatial resolution. In this work, we consider to utilize GAN framework to realize SR on intensity resolution, which is orthogonal to the spatial resolution, realizing image de-quantization.

Compared to the traditional de-quantization methods [4-15], which have been introduced in Section 1, focusing on color gradation smoothing, noise eliminating and banding artifacts reducing. This research proposes an end-to-end learning de-quantization model DAGAN to realize photo-realistic texture details recovering, simultaneously. To the best of our knowledge, this is the first attempt to apply GAN framework for image de-quantization.

## III. ALGORITHMIC PIPELINE

### A. *Problem formulation and constraints*

This research proposes DAGAN model to realize image photo-realistic de-quantization, directly learning an end-to-end mapping between LBD and HBD images, performing LSBs restoring with vivid texture details recovering.

The SR process for the intensity resolution of pixels in image is performed in Fig. 3. Through DAGAN model, the reconstructed pixel intensity is formulated via the end-to-end mapping procedure, considering whole and local contents' relationship, realizing intensity range fulfillment and local value smooth. In this way, LSBs and useful information in LBD image are restored, generating photo-realistic HBD image.

### B. *Network architecture*

GAN [16] architecture transforms the input to a high-dimensional latent representation that is invariant to the specific corruption and then reconstructs the full-dimensional data. We use the same concept for the generating of HBD images.

However, most GAN-based models [21], [22] for image generation are built using convolutional layers. Convolution processes the information in the local neighborhood, thus using convolutional layers alone is computationally inefficient for modeling long-range dependencies in images. In this way, the valuable discriminative features, which usually be regions with abundant edges and detailed texture, are not optimal utilized. Moreover, LBD image is full of massive low-frequency information, thus, the most important task for photo-realistic image de-quantization is to recover the useful high-frequency information, constructing vivid HBD image. Thus, special GAN framework for this task is worth researching.

Fig. 4 gives an overview of the proposed DAGAN scheme. Our DAGAN scheme is consisted of two phases: generative model (G) and discriminative phase (D). G maps the LBD image into the representation feature maps that can be used for vivid texture details recovering, generating corresponding HBD image. The function of D is to classify the given HBD image as real or false image class.

#### *1) Generative network*

According to the image-to-image translation research [17-20], U-net [54] is an effective framework applied in G. Based on the U-net structure, this research adopts the upscale and downscale modules, along with long skip connections between them, realizing multi-scale feature fusion, to construct the basic structure of G network. Under this setting, G can comprehensively holds discriminative contextual as well as useful textural information, resulting in dislodging artifacts and padding textures.

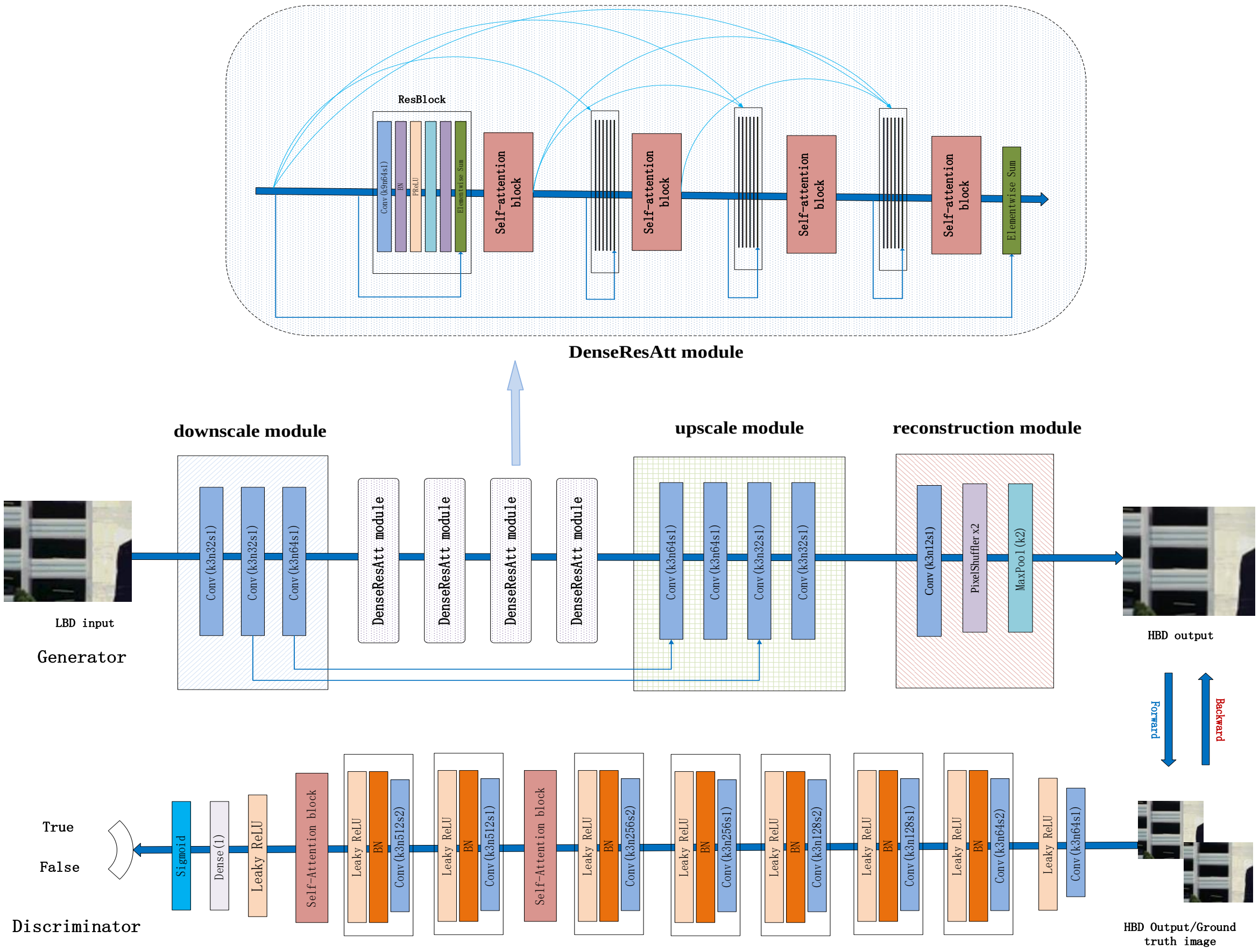

Fig. 4. Architecture of DAGAN with kernel size (k), number of feature maps (n) and stride (s) indicated for corresponding convolutional layer.

As shown in Fig. 4, G in DAGAN model is consisted of four parts: ① upscale module; ② DenseResAtt modules; ③ downscale module; ④ reconstruction module.

Set $I^{LBD}$ and $I^{HBD}$ as the input and output images of DAGAN. Motivated by the structure of U-net [54], in upscale module, we utilize three convolution blocks to extract coarse feature $I_{COARSE}$.

$$I_{coarse} = F_{COA}(I^{LBD}) \quad (1)$$

Where $F_{COA}$(.) donates the upscale module.

$I_{COARSE}$ is then transferred into the sequential DenseResAtt modules, modeling the deep feature $I_{deep}$.

$$I_{deep} = F_{DENSE}(I_{coarse}) \quad (2)$$

Where $F_{DENSE}$ (.) donates the DenseResAtt modules.

Then comes the downscale module, which is consisted by four convolution blocks to convert the deep feature $I_{deep}$ into the fine feature $I_{fine}$.

$$I_{fine} = F_{FINE}(I_{deep}) \quad (3)$$

In Eq.(3) $F_{FINE}$(.) donates the downscale module.

Finally, the fine feature $I_{fine}$ is fed into the final reconstruction module, including the convolution, trained sub-pixel convolution as well as max pooling layers proposed by Shi et al. [48].

$$I^{HBD} = F_{REC}(I_{fine}) \quad (4)$$

Where $F_{REC}$(.) donates the reconstruction module.

*2) DenseResAtt module*

This section performs the details about the proposed DenseResAtt module in Fig. 4.

DenseResAtt module is consisted of four ResAtt groups, constructed by residual block armed with self-attention mechanism. For fully utilize of features and carefully avoid vanishing gradient, each ResAtt group is concentrated to the subsequent groups. Meanwhile, long and short skip connections operate among the DenseResAtt module. Such dense structure allows the G to model deep feature maps which are attentive to the useful feature with abundant edges and detailed texture, conducing to reconstructing the photo-realistic HBD image.

The output $H_i$ of the i th ResAtt group and the input $I_i$ of m th ResAtt group can be modeled below.

$$H_i = \delta_i(\emptyset_i(I_i, W_i) + I_i) \quad (5)$$

$$I_i = [H_1, H_2, H_3, \ldots, H_{i-1}] \quad (6)$$

In Eq.(5), the $I_i$ is the input of i th ResAtt group. $\emptyset$ represents the function of Residual block. δi represents the function of ith attention module. $W_i$ represents the weight set modeled in the i th ResAtt group. For brevity, bias is not performed. Eq.(6) indicates that the input of i th ResAtt group $I_i$ is the concentrate of the former outputs.

In addition, aiming at optimizing the training of DenseResAtt module, the long skip connection is utilized in the module. The functions of long skip connection can be divided into two parts. One is to ease the useful high-frequency information distorting across residual groups. The other one is to promote the network to bypass massive low-frequency information, concentrating on the recovering of high-frequency information.

Aim at modeling the deep feature $I_{deep}$ to learn more useful information of input features, this research realizes the series connection of sequential DenseRes blocks, get better photo-realistic performance. In this research, according to Fig. 5, the number of DenseResAtt modules in DAGAN is set to 4. In this way, the deep feature $I_{deep}$ is formulated.

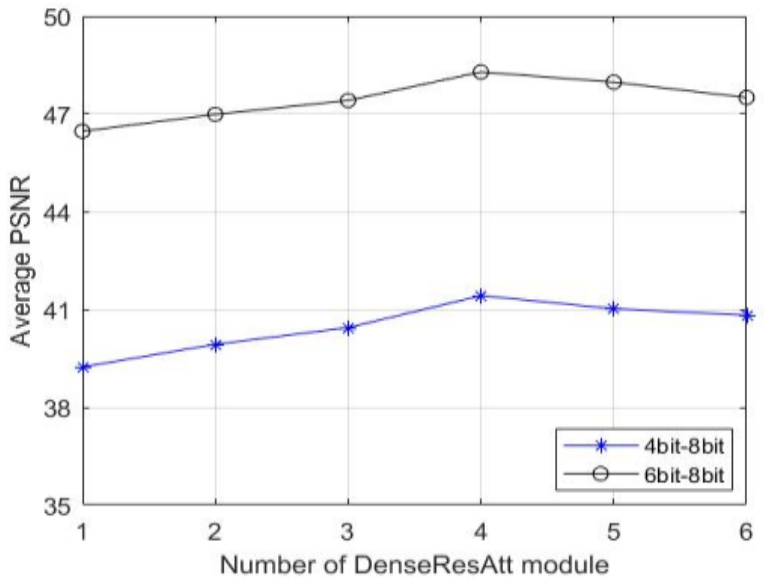

Fig. 5. Performance for different number of DenseResAtt modules. Average PSNRs are calculated on Set 5.

Based on the above settings, sequential DenseResAtt modules can recover useful features via the discriminative learning of coarse level residual information.

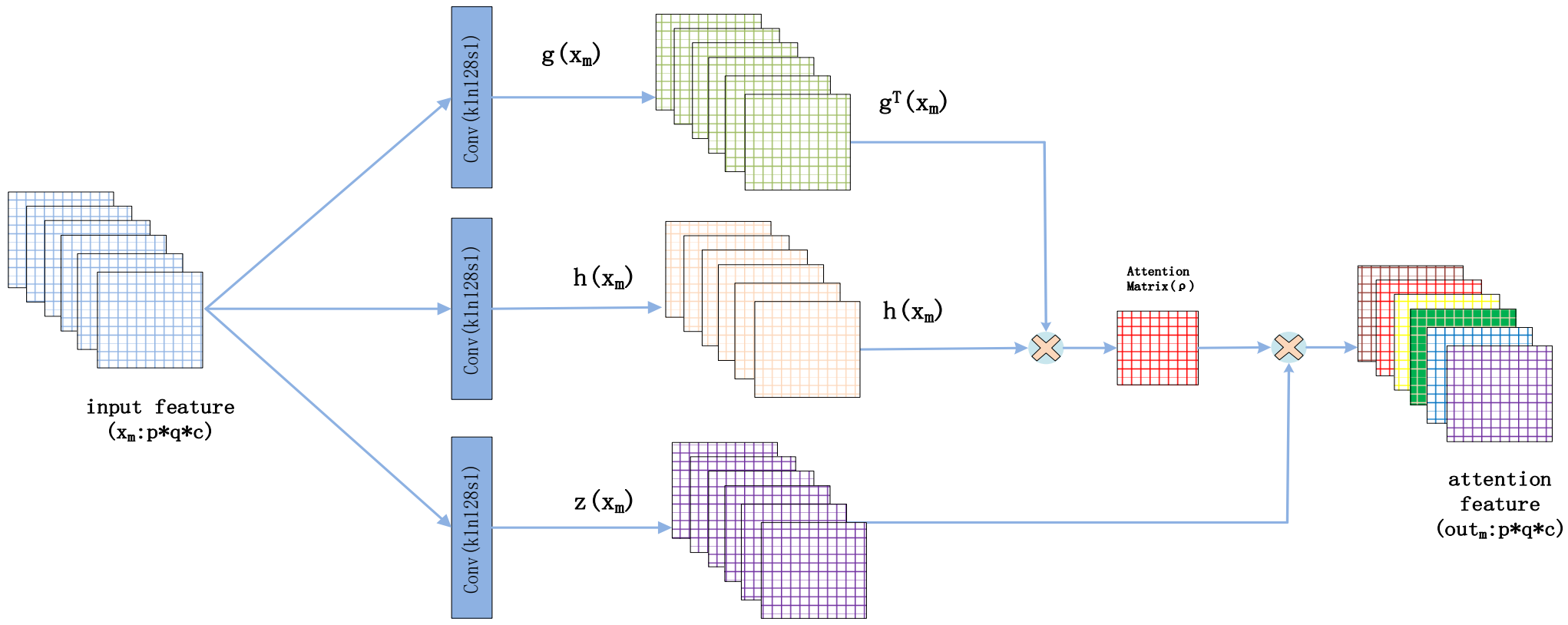

Fig. 6. Architecture of self-attention block

.

### 3) *Self-attention block*

To make a further step towards discriminative features learning, we adopt the self-attention block based on the non-local model [59, 52], enabling both the G and the D to efficiently model the relationships between local area and widely spatial regions. In this way, long-range dependencies in deep feature $F_{deep}$, which is modeled by sequential DenseResAtt modules, is constructed. Fig. 6 illustrates the architecture of self-attention block.

Firstly, the input features from the former residual block, $x_m \in R^{p*q*c}$, are processed by two 1*1 convolution layers, respectively, getting two corresponding feature space g, h, forming the attention matrix $\rho_{j,i}$, which represents the extent to which the model towards the $i^{th}$ location when constructing the $j^{th}$ area.

$$g(x_m) = W_g x_m \tag{7}$$

$$h(x_m) = W_h x_m \tag{8}$$

$$\rho_{j,i} = \frac{\exp\left[\left(g(x_m)(i)\right)^T (h(x_m)(j))\right]}{\sum_{i=1}^{q} \exp\left[\left(g(x_m)(i)\right)^T (h(x_m)(j))\right]} \tag{9}$$

Then the output attention feature, $out_m \in R^{p*q}$, is formulated, where

$$z(x_m) = W_z x_m \tag{10}$$

$$out_m = \sum_{i=1}^{q} \rho_{j,i} z(x_m(i)) \tag{11}$$

In Eq.(7)-Eq.(11), $W_g \in R^{a*p}$, $W_h \in R^{a*p}$, $W_h \in R^{p*p}$ are the weight matrics after modeling. In addition, a is set to p/8 in this research.

Moreover, aiming at making more optimal use of existing details, we combine the input feature $x_m$ and the output attention feature $out_m$, forming the final result.

$$y_m = \sigma out_m + x_m \tag{12}$$

In Eq.(12), σ is set to zero initially and adjusted in the followed learning progress.

Under this setting, discriminative details can be captured based on cues from widely spatial regions.

### 4) *Discriminator network*

According to recent works [23,56,57,58], the D network is useful for preventing over-smoothed images. Thus, we utilize the D to distinguish the fool of G, making a distinction between the reconstructed HBD images and ground truth HBD images. The architecture of D is shown in Fig. 4.

The architecture of the D includes staked convolutional layers with embedded self-attention blocks. The feature dimension in staked convolutional layers is enlarged by the fixed constant 2 from 64 to 512, according to VGG model [59]. The fully connected layers is utilized from [60]. The embedded self-attention blocks promote G to focus on the discriminative content of the input image.

The discriminative loss, also named as adversarial loss, is back-propagated during the training process to update G network. Through this manner, we can promote G to generate vivid HBD images, as shown in Fig. 2. Note that D is only utilized in the training stage.

## C. Loss function

### 1) Content loss

In this research, a special content loss $l^{\mathrm{HBD}}$ is designed for this quality enhancement task. $l^{\mathrm{HBD}}$ is consisted of several loss components with different weighted, learning distinct desirable features to recover image content, color, and texture.

According to previous research, pixel-wise Mean Square Error (MSE) loss is commonly used in image restoration as well as SR methods. Thus, $l_{MSE}^{\mathrm{HBD}}$ (MES loss) is utilized as part of our content loss $l^{\mathrm{HBD}}$. The MSE loss is calculated in Eq.(13).

$$l_{MSE}^{\mathrm{HBD}} = \frac{1}{WH}\sum_{x=1}^{W}\sum_{y=1}^{H}(I_{x,y}^{\mathrm{HBD}} - G_{\theta_G}(I^{\mathrm{LBD}})_{x,y})^2 \quad (13)$$

Where $I^{\mathrm{LBD}}$ is the input LBD image, $I^{\mathrm{HBD}}$ represents the ground truth HBD image. W and H are the width and height of the image. Admittedly, MES loss is conducive to getting high PSNR. Nevertheless, high frequency features are always cannot be fully utilized only depend on MSE feed-forward optimization, resulting in perceptually shortage with overly smooth textures. Thus, perceptual loss [61] is added to this research.

Due to diverse image quality, the global visual percept of LBD image and HBD image is different. Aiming at performing high quality visual images, we also constrain the HBD images to share the same high frequency contents as their ground truths. Thus, based on the perceptual loss proposed by the John-son et al. [61], which extracts the feature maps of high-level features of the reconstructed HBD image and the corresponding ground-truth image, via utilizing the classical VGG network [59] with pre-trained parameters, this research proposes the new perceptual loss $l_{VGG}^{\mathrm{HBD}}$ to evaluate perceptually characteristics.

$l_{VGG}^{\mathrm{HBD}}$ loss in Eq.(14) is designed upon pre-trained VGG model. In Eq.(14), $\alpha_{i;j}$ represents the feature maps modeled by the j-th convolution layer, which is located between the followed activation layer and former max pooling layer. $W_{i;j}$ and $H_{i;j}$ represent the dimensions of the feature maps.

$$l_{VGG}^{\mathrm{HBD}} = \frac{1}{\mathrm{W}_{i,j}\mathrm{H}_{i,j}}\sum_{x=1}^{W}\sum_{y=1}^{H}(\alpha_{i,j}(I^{\mathrm{LBD}})_{x,y} - \alpha_{i,j}(G_{\theta_G}(I^{\mathrm{LBD}})_{x,y})^2 \quad (14)$$

In all, the specified content loss $l^{\mathrm{HBD}}$ in this task is the weighted sum of MSE loss ($l_{MSE}^{\mathrm{HBD}}$) and VGG loss ($l_{VGG}^{\mathrm{HBD}}$). In this research, β is set to 0.5.

$$l^{HBD} = \beta l_{MSE}^{\mathrm{HBD}} + (1-\beta) l_{VGG}^{\mathrm{HBD}} \quad (15)$$

### 2) Adversarial loss

In GAN [16] framework, G captures the data distribution from input training images $I^{\mathrm{LBD}}$ and $I^{\mathrm{HBD}}$, generating the synthetic images $G(I^{\mathrm{LBD}})$ to fool D. D estimates the image to judge whether it is the real image in training data or the synthetic product $G(I^{\mathrm{LBD}})$ from G. The adversarial loss functions to describe this game are as follows:

$$l_{\mathrm{adversarial}} = min_\theta max_\rho E_{I^{HBD}\sim p_{train}(I^{HBD})}\left[\log D_\rho(I^{HBD})\right] + E_{I^{LBD}\sim p_G(I^{LBD})}\left[\log\left(1 - D_\rho\left(G_\theta(I^{LBD})\right)\right)\right] \quad (16)$$

$$l_{Gen} = E_{I^{LBD}\sim p_G(I^{LBD})}\left[\log\left(1 - D_\rho\left(G_\theta(I^{LBD})\right)\right)\right] \quad (17)$$

In Eq.(16), θ and ρ are the parameters in G and D, which are obtained in the optimizing progress. Here, $D\left(G(I^{LBD})\right)$ is the probability that the reconstructed image $G(I^{LBD})$ is a real high-quality image.

This adversarial processing encourages perceptually superior solutions resided in the subspace, the manifold, of natural images. This is in contrast to image restoration as well as SR solutions obtained by minimizing pixel-wise error measurements, such as the MSE. Moreover, DAGAN scheme adopts the attention mechanism to G and D, which are trained in an alternating fashion by minimizing the hinge version of the adversarial loss.

### 3) Content loss

Above all, the performance of our DAGAN model is depended on the loss function $l_{Dis}^{HBD}$ and $l_{Gen}^{HBD}$, which are illustrated as follows.

The loss of G, showing in Eq.(18), is consisted of content loss as well as adversarial loss. This specified loss promotes our model to generate HBD results based on the manifold of natural images, aiming at fooling the D network. The loss $l_{Gen}^{HBD}$ is defined as follows.

$$l_{Gen}^{HBD} = \underbrace{l^{HBD}}_{content\ loss} + \underbrace{10^{-2} l_{Gen}}_{adversarial\ loss}$$

$$= \underbrace{\beta l_{MSE}^{\mathrm{HBD}}}_{for\ MSE\ losses} + \underbrace{(1-\beta) l_{VGG}^{\mathrm{HBD}}}_{for\ VGG\ losses} + \underbrace{10^{-2} l_{Gen}}_{adversarial\ loss} \quad (18)$$

$$l_{Dis}^{HBD} = l_{\mathrm{adversarial}} \quad (19)$$

The loss of D is represented in Eq.(19). In this way, DAGAN is trained by gradient descending the G's loss $l_{Gen}^{HBD}$ as well as D's loss $l_{Dis}^{HBD}$ to achieve higher visual quality and structural similarity to the original ground truth HBD images.

## IV. Experiments

## A. Experiment settings and database

### 1) Experiment setting

In this research, the famous DIV2K database is utilized as training set. We randomly cropped 64*64 images blocks from the 800 images in the database, forming the training dataset. Note that these image blocks are flipped and rotated with 90°, 180°, and 270° to enlarge the training dataset.

This research creates LBD images by reducing the bit-depth of original HBD images. Due to most of current monitors are with 8-bit, we quantizes the images in training set to 4-bit and 6-bit and then enhances them back to 8-bit through comparable approaches.

This network was trained 80 epoches, taking about 2 hours for each epoch on a Nvidia Titan X GPU. The learning rate is set to 10-4 at beginning, reducing by a factor of 10 in the following every 10 epochs. Adam optimizer is adopted in this research.

### 2) Testing database description

To evaluate the efficiency of the DAGAN, we conduct extensive testing experiments. For more convincing testing, according to the settings in traditional SR and BDE methods, public testing databases are introduced, including Set5 dataset [63], 6 images from USC-SIPI dataset [64], 100 images from BSD100 dataset [65] and 100 images from VOC2012 dataset [62] are chose.

*3) Comparable approach*

**Baseline:** We only trained the G in DAGAN, without D, forming DA-net as baseline.

**The proposed approach:** DAGAN

**De-quantization approach:** ZP [4], MIG [5], MRC [7], IPAD [10], ICH-CNN [13] as well as BDEN [12].

## B. *Subjective evaluation*

The subjective results of smooth transition as well as edge areas in testing databases are performed in Fig. 7 and Fig. 8. According to the visual comparison of different methods on smooth transition areas, where false contour artifacts are the common noises to appear, showing in Fig. 7, the conclusions can be get as follows. First, ZP, MIG and MRC methods result in obvious false contour artifacts, just like the contours among the 'sky', which have been enlarged and shown in Fig. 7 (b) (c) and (d). The reason to this phenomenon is that pixel-based algorithms ZP, MIG and MRC are the traditional de-quantize methods without taking local context and local consistency into consideration, resulting in ignoring the color transition trend. Second, although IPAD takes the local neighborhood into consideration to reduce most banding artifacts around continuous contours. However, its block-wise handling principle weakens the ability to reconstruct globally smooth areas, resulting in some isolated noises existing in smooth transition areas, performed in Fig. 7 (e). Third, the inverse half-toning method ICH-CNN overlooks the texture recover ability, thus, it also blurs the textural details, performed in Fig. 7 (f). Fourth, the BDEN utilizes a local adaptive adjustment technique to eliminate perceptible artifacts in flat area, refines the visual quality of smooth transition area, performed in Fig. 7 (g). Compared to these methods, the contours among the 'sky' are well handled with smooth transition to their neighborhood areas by the DAGAN method, performed in Fig. 7 (h). This can be attributed to the attentive structure of DAGAN, which is equipped with superior discriminative learning ability, modeling long-range dependencies across image regions.

Then, Fig. 8 shows the close-ups of edges of objects after the processing of these methods. Broken or reshaped structures are the common challenging problems in processing procedures. Owning to the context independent property, the pixel-based algorithms ZP, MIG and MRC cannot produce sharper edge and divide the boundaries of wall and pepper from the background clearly. Thus, unnatural artifacts is adhered to the edges of wall and pepper, showing in Fig. 8 (b) (c) and (d). Via the enhancement processes on both local and global views in IPAD method, false contour artifacts is well handled, however, the boundaries are still mixed with background, performed in Fig. 8 (e). According to Fig. 8 (f) and (g), although ICH-CNN and BDEN can recover relative shape edge, they still cannot avoid producing blurs adhered to backgrounds. This is due to the shortage of their traditional convolution layer with local receptive field, they cannot learn global dependencies in the input image, resulting in unsatisfied overall visual effect. Owing to the effective information restore function of GAN network, the proposed method DAGAN can recover shape edges without introducing over-blur areas or false contour artifacts, showing in Fig. 8 (h). Two reasons can explain for this phenomenon, one is that the information restore function of GAN network is effective, another one is that the deep attentive structure of G enables the DAGAN model to detect the most discriminative features.

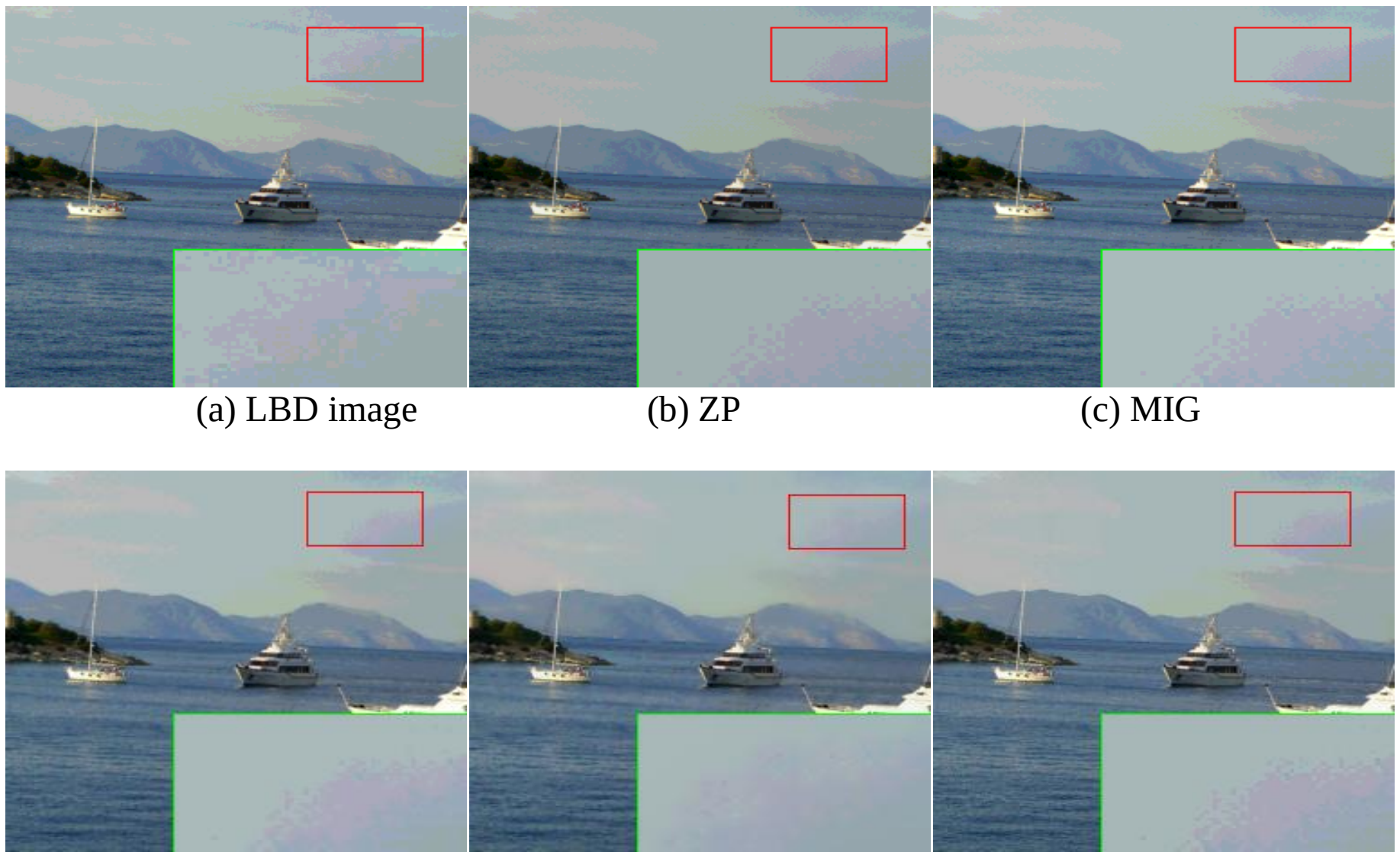

(a) LBD image (b) ZP (c) MIG

(d) MRC (e) IPAD (f) ICH-CNN

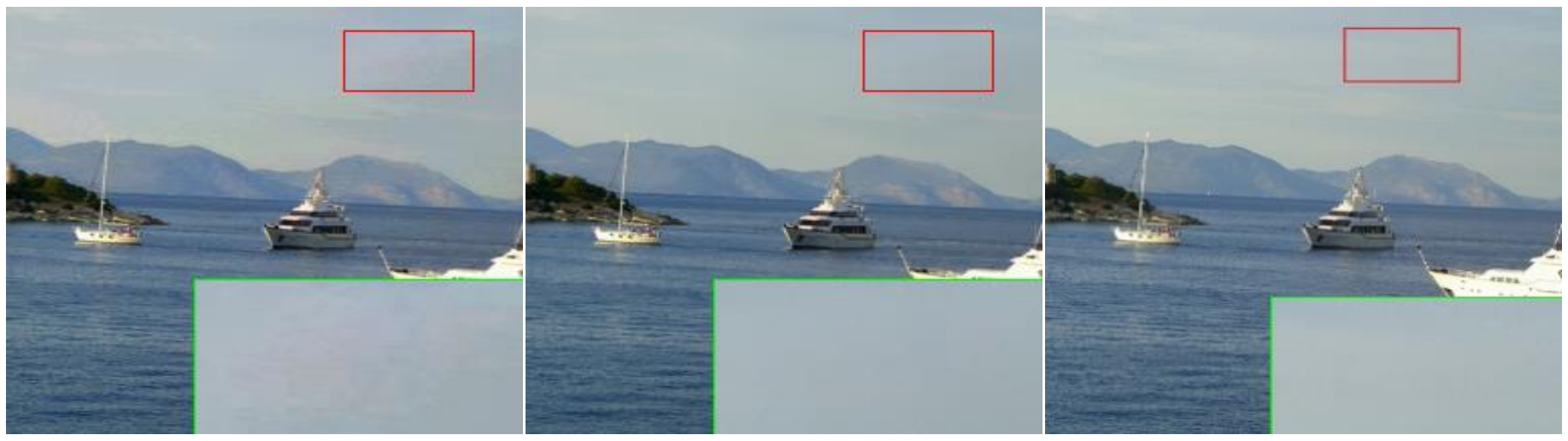

(g)BDEN (h) DAGAN (i) Ground Truth

Fig. 7. Visual comparison of overall effect and smooth transition area of scenery image from VOC2012 dataset with de-quantization methods (4-bit to 8-bit). Additionally the enlarged smooth transition area of each image is performed at the bottom for better comparison.

Moreover, Fig. 9 performs the close-ups of facial components and texture areas after the processing of these methods on the images in testing datasets. Although not perfect, our method can reconstruct more detailed intensity range as well as photo-realistic content than the existing methods. While there is amplified noise in other results, our results exhibit photo-realistic texture reproduction without introducing over-blur parts or false contour artifacts. Please enlarge images in the electronic version.

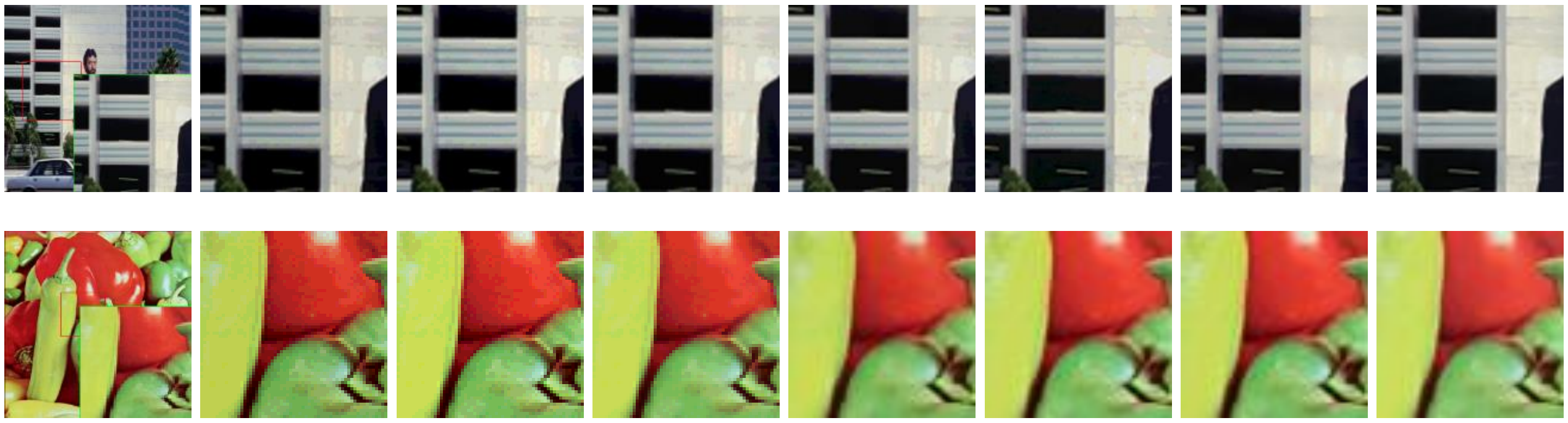

(a) Ground Truth (b) ZP (c) MIG (d) MRC (e) IPAD (f) ICH-CNN (g) BDEN (h) DAGAN

Fig. 8. Visual comparisons of compared de-quantization methods (4-bit to 8-bit) on texture and edge areas for image ‘119082’ in BSD 100 database and ‘Pepper’ in in USC-SIPI database.

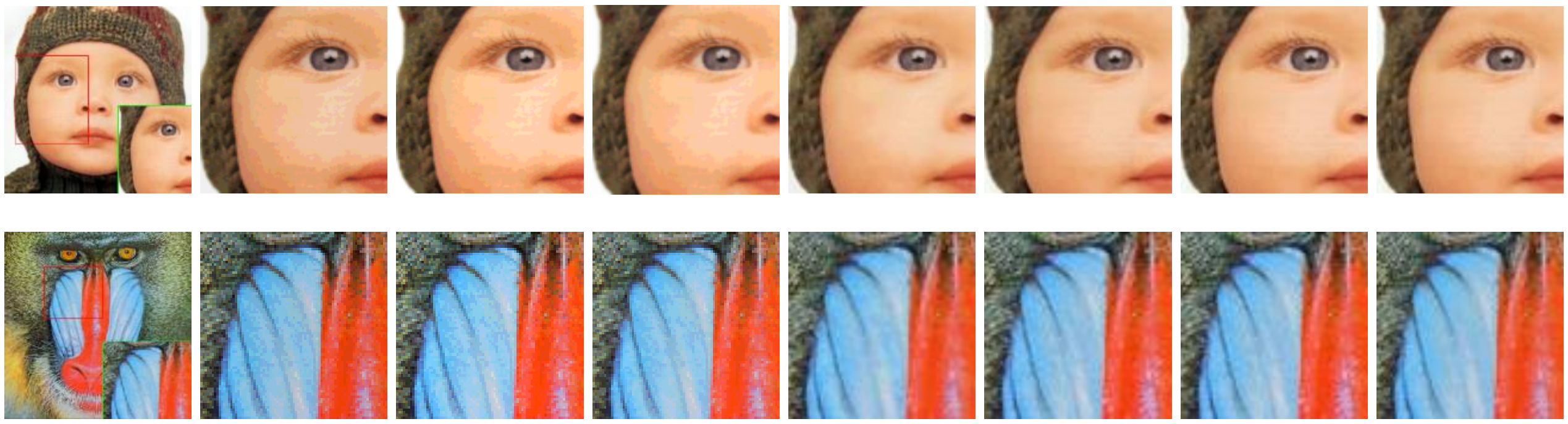

(a) Ground Truth (b) ZP (c) MIG (d) MRC (e) IPAD (f) ICH-CNN (g) BDEN (h) DAGAN

Fig. 9. Visual comparisons of compared de-quantization methods (4-bit to 8-bit) on facial components and texture areas for image ‘Baby’ in Set 5 and ‘Baboon’ in USC-SIPI database.

### C. *Objective evaluation*

According to the conclusion of SRGAN [23], PSNR cannot capture perceptually relevant differences, such as vivid texture detail. Thus, the SSIM index is introduced to our evaluation work. In this way, for objective evaluation, we calculate the PSNR and SSIM of the experiment results with ground truth images, showing in Table I, II and III. What should be noticed is that, we measure the performance by the average PSNR and the structural SSIM scores on each entire testing dataset.

TABLE I
AVERAGE PSNR [dB] AND SSIM VALUES OF COMPARED DE-QUANTIZATION METHODS ON BSD100 DATASET (HBD=8-bit).

| | PSNR | SSIM | PSNR | SSIM |
|---|---|---|---|---|
| | LBD=4 bit | | LBD=6 bit | |
| ZP [4] | 27.9156 | 0.9530 | 39.3518 | 0.9661 |
| MIG [5] | 33.2123 | 0.9518 | 45.8629 | 0.9684 |
| MRC [7] | 35.8386 | 0.9641 | 45.9431 | 0.9738 |
| IPAD [10] | 37.7635 | 0.9726 | 46.1774 | 0.9795 |
| ICH-CNN [13] | 35.1535 | 0.9590 | 45.9425 | 0.9633 |
| BDEN | 38.8200 | 0.9792 | 47.7881 | 0.9859 |
| DA-net | 39.2750 | 0.9806 | 48.5629 | 0.9913 |
| DAGAN | **40.1271** | **0.9830** | **49.2836** | **0.9952** |

As can be observed from Table I, on BSD100 dataset, ZP and MIG are with lower PSNR and SSIM since they ignore the critical local contexts. MRC and IPAD predict the image from the statistical aspect of visual, thus, they produce natural images with irregular textures, resulting in higher PSNR/SSIM. Due to the shortage of convolution network with local receptive field, ICH-CNN cannot learn long-term dependencies in the input image, resulting in low PSNR and SSIM. According to the result, advantage of the BDEN is more obvious. Due to its deliberate design for inferring photo-realistic natural images, it get large-scale increase than the previous methods, more than 1.9 dB compared to MRC method in 6-bit to 8-bit procedure on PSNR. Compared with these competitive algorithms, the proposed DAGAN algorithm is more effective, achieving best performance on both PSNR and SSIM. The average increases over the second method (DA-net) achieve as high as 0.85 dB and 0.024 in PSNR and SSIM, respectively, in 4-bit to 8-bit procedure. We attribute the remarkable performance increase to the efficiency of the proposed sequential DenseResAtt modules, which vests G the ability to study the long-range dependencies in the feature maps, approximating the original image distribution preciously and forming HBD image with vivid high-frequency information and smooth transition background.

TABLE 2. PSNR [dB] AND SSIM VALUES OF COMPARED DE-QUANTIZATION METHODS (4-bit TO 8-bit) FOR ON IMAGES BABY, FEMALE, JELLY BEANS, SPLASH AND TIGER IN TESTING SETS.

| IMAGE | ZP [4] | | MIG [5] | | MRC [7] | | IPAD [10] | | ICH-CNN [13] | | BDEN [12] | | DA-net | | DAGAN | |
|---|---|---|---|---|---|---|---|---|---|---|---|---|---|---|---|---|
| | PSNR | SSIM | PSNR | PSNR | PSNR | SSIM | PSNR | SSIM | PSNR | SSIM | PSNR | SSIM | PSNR | SSIM | PSNR | SSIM |
| Baby | 31.5917 | 0.9505 | 34.4645 | 0.9497 | 39.1895 | 0.9660 | 39.1720 | 0.9667 | 33.0347 | 0.9681 | 39.9774 | 0.9774 | 39.0967 | 0.9649 | **40.0783** | **0.9808** |
| Female | 31.1427 | 0.9764 | 34.4595 | 0.9743 | 36.9214 | 0.9787 | 41.2177 | 0.9904 | 37.9276 | 0.9878 | 39.7442 | 0.9711 | 39.7927 | 0.9900 | **41.2004** | **0.9925** |
| Jelly beans | 31.4164 | 0.9575 | 35.1855 | 0.9545 | 38.6006 | 0.9697 | 39.8491 | 0.9780 | 35.5343 | 0.9836 | 36.3353 | 0.9834 | 40.6896 | 0.9834 | **42.9426** | **0.9845** |
| Splash | 31.3518 | 0.9436 | 36.1167 | 0.9392 | 39.3643 | 0.9587 | 39.5234 | 0.9662 | 34.0782 | 0.9637 | 38.3256 | 0.9710 | 38.4963 | 0.9627 | **40.3772** | **0.9695** |
| Tiger | 31.0759 | 0.9735 | 34.6533 | 0.9733 | 39.5546 | 0.9709 | 38.5542 | 0.9775 | 36.3393 | 0.9785 | 41.3921 | 0.9876 | 40.0601 | 0.9842 | **42.7127** | **0.9891** |
| Baboon | 31.1942 | 0.9724 | 35.1347 | 0.9711 | 38.9166 | 0.9782 | 38.4247 | 0.9758 | 36.6367 | 0.9706 | 39.9210 | 0.9805 | 40.2485 | 0.9827 | **41.3837** | **0.9886** |
| Average | 31.2954 | 0.9623 | 35.0023 | 0.9603 | 38.7578 | 0.97036 | 39.4568 | 0.9757 | 35.5918 | 0.9753 | 39.2826 | 0.9785 | 39.73065 | 0.9779 | **41.4491** | **0.9841** |

TABLE III
AVERAGE PSNR [dB] AND SSIM VALUES OF COMPARED DE-QUANTIZATION METHODS (4-bit TO 8-bit) ON SET5 AND VOC2012 DATASET.

| | Set5 | | VOC2012 | |
|---|---|---|---|---|
| | PSNR | SSIM | PSNR | SSIM |
| ZP [4] | 31.6770 | 0.9596 | 31.1942 | 0.9724 |
| MIG [5] | 35.2285 | 0.9574 | 37.9674 | 0.9719 |
| MRC [7] | 37.0948 | 0.9730 | 37.9166 | 0.9782 |
| IPAD [10] | 38.0148 | 0.9734 | 38.4247 | 0.9758 |
| ICH-CNN [13] | 38.1385 | 0.9761 | 36.6258 | 0.9689 |
| BDEN | 39.2455 | 0.9785 | 38.9727 | 0.9753 |
| DA-net | 40.6411 | 0.9810 | 39.2831 | 0.9819 |
| DAGAN | **41.4223** | **0.9826** | **40.5436** | **0.9838** |

Moreover, the PSNR/SSIM results of compared de-quantization methods for the images in USC-SIPI [64] database are performed in Table II. At the same time, the average PSNR/SSIM results on the other public databases, Set5 [64] and VOC2012 [62] databases, are performed in Table. 3. All the experiment results prove the superiority of DAGAN algorithm.

According to Table II and Table III, context-independent algorithms such as ZP and MIG produce images with low quality, resulting in unsatisfied PSNR/SSIM. Even though the MRC and IPAD tend to generate better results on PSNR and SSIM, however, they still cannot avoid producing false contour artifacts, which have been performed in column (d) and (e) in Fig. 8 and 9. Via comparation, BDEN can effectively eliminate the perceptible artifacts in flat area, such as the performed images in column (g) in Fig. 7. However, when it comes to the ability for reconstructing the high quality details, BDEN performs unsatisfied performance, just like column (g)s in Fig. 8 and 9 show. Thus, the PSNR/SSIM of BDEN still cannot achieve the best performance. Compared with these state-of-the-art algorithms, the proposed DAGAN algorithm shows excellent processing ability for not only photo-realistic texture recovering but also greatly suppressed false contours and chroma distortions. Thus, it is widely adapted to both significant gradient diversity and smooth gradient diversity images, getting more excellent results (shown in Table. II and Table. III) and high subjective visual quality (performed in Fig. 7, 8 and 9) on all testing databases. The average gains over the second algorithm DA-net achieve as high as 0.78 dB and 0.016 in PSNR and SSIM on Set 5 database, respectively.

### *D. Evaluation for SR on image intensity*

This research proposes DAGAN framework to realize SR on intensity resolution, which is orthogonal to the spatial resolution, realizing image de-quantization. Thus, a subjective evaluation for the SR performance of DAGAN is implemented.

Fig. 10 performs the intensity comparation of the pixels in image ‘Jelly’. According to the result, DAGAN model well considers the whole and local contents’ relationship, realizing intensity range fulfillment and local values smooth. In this way, photo-realistic image de-quantization is realized.

Moreover, the intensity distributions of image ‘Jelly’ in corresponding LBD, DAGAN and ground truth HBD images are performed in Fig.11. This also proves that the intensity distribution of the generate HBD image and ground truth HBD image are similar. Thus, DAGAN can well realize SR on the intensity resolution, realizing photo-realistic image de-quantization.

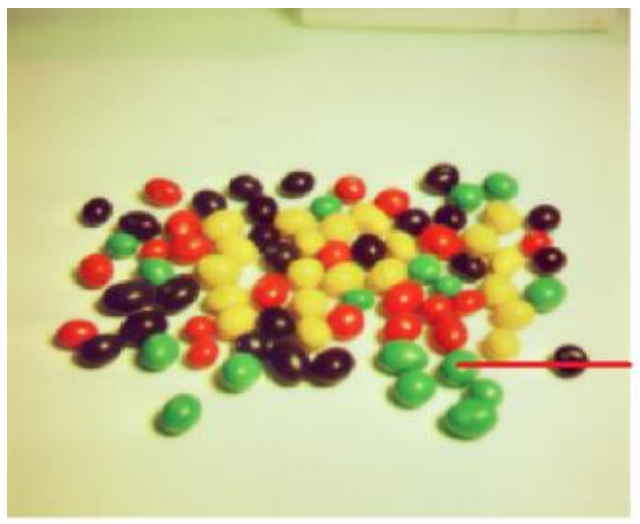

(a) Comparision pixels on a row in the image ‘Jelly’

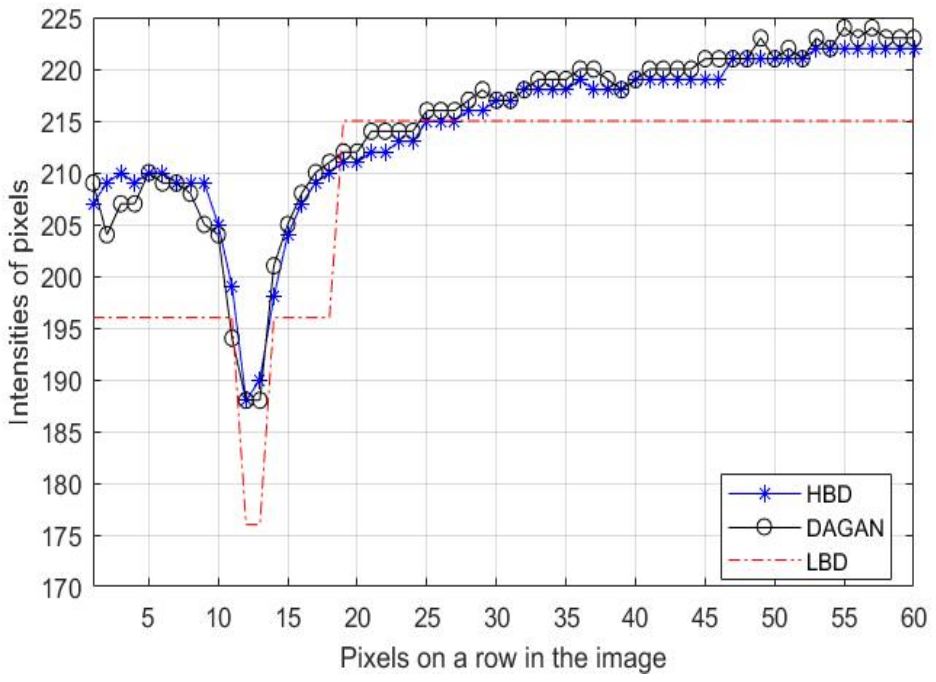

(b) Intensity of pixels

Fig. 10. Intensity comparation of the pixels in corresponding ground truth HBD, DAGAN and LBD images.

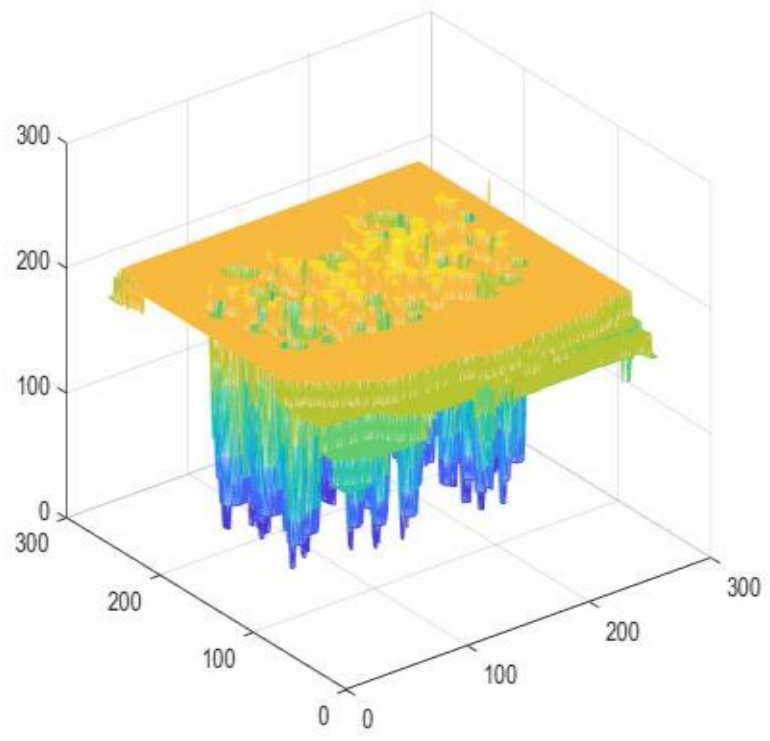

(a) Distribution of intensity in LBD image.

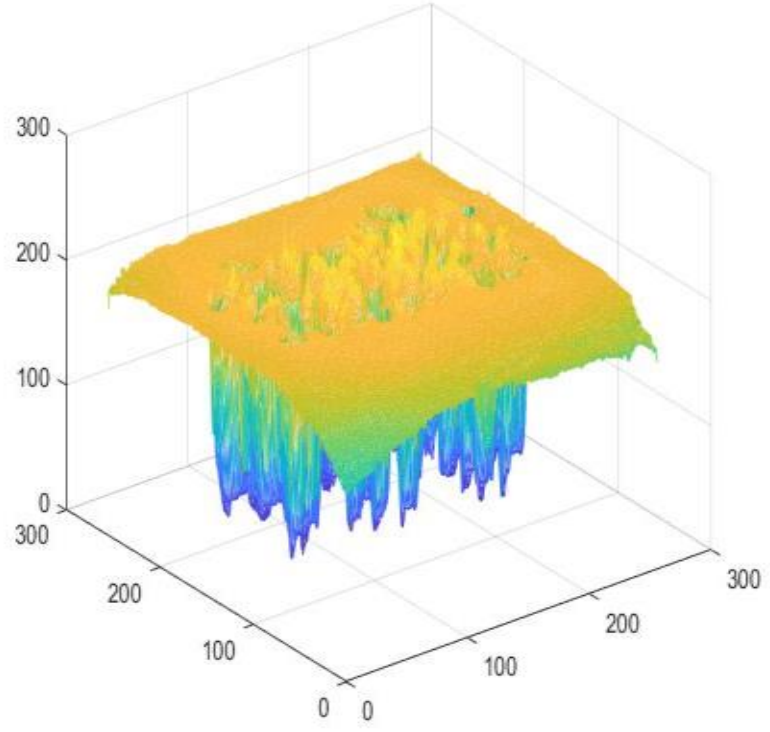

(b) Distribution of intensity in HBD image (generated by DAGAN).

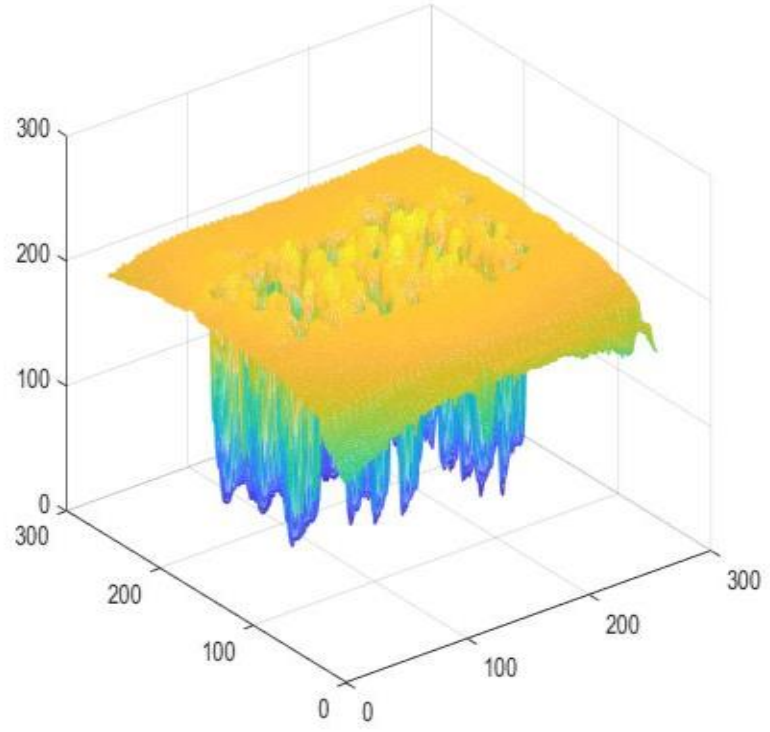

(c) Distribution of intensity in ground truth HBD image.

Fig. 11. Intensity distribution in corresponding LBD, generated and ground truth HBD images.

### *E. Ablation Analysis*

**Effectiveness of Attention Mechanism:** As shown in Fig.12 (b), we can observe that the visual performance of DAGAN* model (DAGAN without self-attention blocks) suffers from texture distortion and blur artifacts. By employing attention mechanism, we can reconstruct HBD image with photo-realistic texture, showing in Fig.12 (e). Not surprisingly, DAGAN achieves higher reconstruction performance, increasing approximately 1.26 dB in PSNR, showing in Tab. IV.

**Effectiveness of Discriminator network:** Aiming at generating photo-realistic HBD images without over-smoothed details, DAGAN employs the adversarial study, including not only generator network but also discriminator network. To show the performance of discriminator network, DA-net model (DAGAN without discriminator network) is introduced for comparison (Fig.12 (c)). As indicated in Table. IV, using the discriminator network, we can improve the quantitative performance by 1.39 dB and 0.0066 in PSNR and SSIM, respectively.

**Loss Functions:** Table. V indicates the influences of various losses on the de-quantization performance. As indicated in Table. V and Fig. 12, only utilize the MSE loss (Fig. 12 (d)) leads to unnatural artifacts and banding effects. Thus, combine with the VGG loss (Fig. 12 (e)) makes the texture and edge more sharper and realistic. Thus, the content loss which is the combination of MSE loss and VGG loss, in DAGAN model, improves the de-quantization results qualitatively and quantitatively, achieving 40.5436 dB and 0.9838 in PSNR and SSMI, respectively.

TABLE IV
ABLATION STUDY OF NETWORK MODULE (ON BSD100 SUBSET)

| | DAGAN* | DA-net | DAGAN |
|---|---|---|---|
| PSNR [dB] | 39.1570 | 39.2831 | 40.5436 |
| SSIM | 0.9772 | 0.9819 | 0.9838 |

TABLE V
ABLATION STUDY ON THE CONTENT LOSS (ON BSD100 SUBSET)

| | VGG loss ($l_{VGG}^{HQ}$) | MSE loss ($l_{MSE}^{HQ}$) | $l_{VGG}^{HQ}+l_{MSE}^{HQ}$ (DAGAN) |
|---|---|---|---|
| PSNR [dB] | 38.5502 | 39.1996 | 40.5436 |
| SSIM | 0.9705 | 0.9826 | 0.9838 |

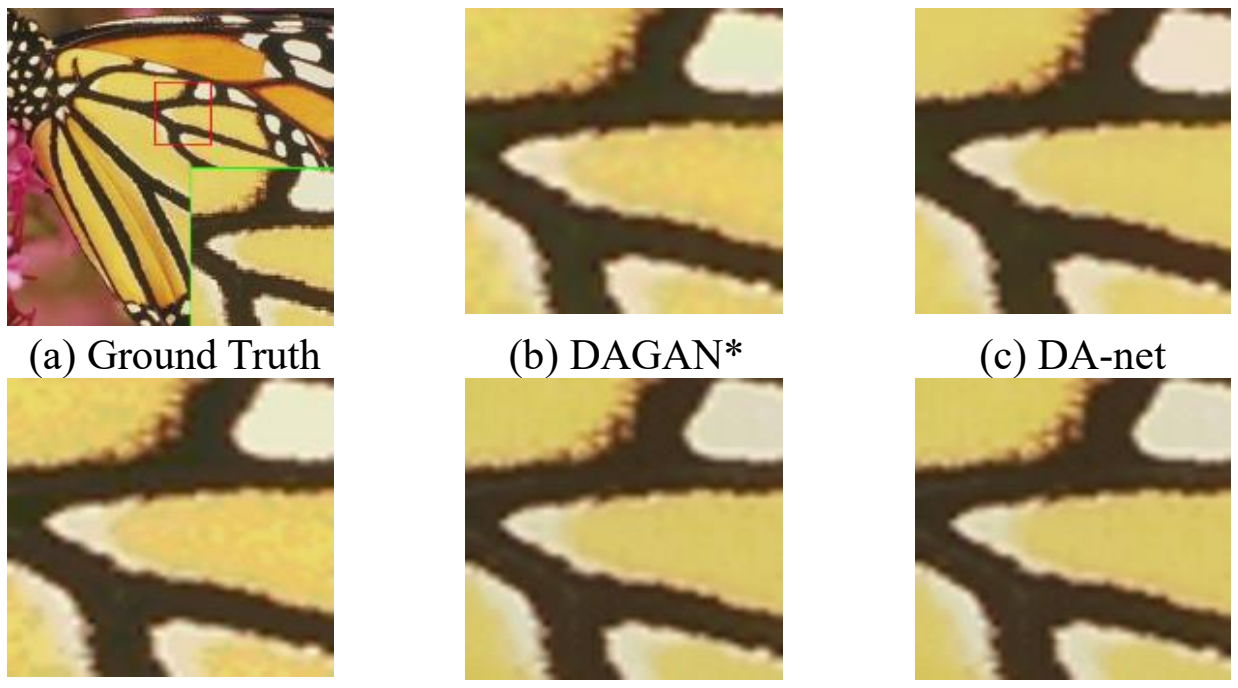
(a) Ground Truth (b) DAGAN* (c) DA-net
(d) MSE loss (e) VGG loss (f) DAGAN
Fig. 12. Visual results of compared ablation analysis methods (4-bit to 8-bit).

## V. CONCLUSION

In this paper, we propose DAGAN model to perform intensity super-resolution, realizing photo-realistic de-quantization. We propose a novel DenseResAtt module, which enables the network to pay attention to useful information. Through this GAN model, the LBD image is represented as effective deep feature maps to reconstruct the vivid HBD image. Until now, this is the first attempt to apply GAN framework for image de-quantization. Comparative trials on Set 5, USC-SIPI, VOC2012, as well as BSD100 databases show that the DAGAN model can perform excellent visual quality as well as satisfied quantitative performance when compared to other state-of-the-art techniques, including ZP, MIG, MRC, IPAD, ICH-CNN, as well as BDEN. In the next research, we will extend this work to more challenge scenarios, such as the de-quantization task with multiple types of artifacts and low-quality video processing [66].

## APPENDIX

Appendixes, if needed, appear before the acknowledgment.

## ACKNOWLEDGMENT

The authors would like to thank the editor and the anonymous reviewers for their valuable comments and constructive suggestions. This work is supported by the National Natural Science Foundation of China (No.61871123 &No.61802203), Key Research and Development Program in Jiangsu Province (No.BE2016739), Natural Science Foundation of Jiangsu Province (No.BK20180311) and a Project Funded by the Priority Academic Program Development of Jiangsu Higher Education Institutions.

## REFERENCES

[1] Mantiuk, Rafal, Grzegorz Krawczyk, Karol Myszkowski, and Hans-Peter Seidel, "Perception-motivated high dynamic range video encoding," *ACM Transactions on Graphics (TOG)*, vol. 23, no. 3, pp. 733-741, 2004.

[2] Sullivan, Gary J., and Jens-Rainer Ohm, "Meeting report of the fourth meeting of the Joint Collaborative Team on Video Coding (JCT-VC), Daegu, KR, 20–28 January 2011," *Document JCTVC-D500*, Daegu, KR, 2011.

[3] Chen, Qiubo, Hengyu Zhao, Hongbin Sun, and Nanning Zheng, "Exploiting bit-depth scaling for quality-scalable energy efficient display processing," in *IEEE International Symposium on Circuits and Systems (ISCAS)*, 2015, pp. 2357-2360.

[4] Wan P, Au O C, Tang K, et al, "From 2d extrapolation to 1d interpolation: Content adaptive image bit-depth expansion," in *IEEE International Conference on Multimedia and Expo*, 2012, pp. 170-175.

[5] Ulichney, Robert A., and Shiufun Cheung, "Pixel bit-depth increase by bit replication," In *Color Imaging: Device-Independent Color, Color Hardcopy, and Graphic Arts III*, International Society for Optics and Photonics, vol. 3300, pp. 232-242, 1998.

[6] Cheng, Cheuk-Hong, Oscar C. Au, Chun-Hung Liu, and Ka-Yue Yip, "Bit-depth expansion by contour region reconstruction," *IEEE International Symposium on Circuits and Systems*, IEEE, 2009, pp. 944-947.

[7] Mittal, Gaurav, Vinit Jakhetiya, Sunil Prasad Jaiswal, Oscar C. Au, Anil Kumar Tiwari, and Dai Wei, "Bit-depth expansion using minimum risk based classification," *Visual Communications and Image Processing*, pp. 1-5, 2012.

[8] Wan, Pengfei, Gene Cheung, Dinei Florencio, Cha Zhang, and Oscar C. Au, "Image Bit-Depth Enhancement via MaximumA PosterioriEstimation of AC Signal," *IEEE Transactions on Image Processing*, no. 6, pp. 2896-2909, 2016.

[9] Liu, Xianming, Gene Cheung, Xiaolin Wu, and Debin Zhao, "Random walk graph Laplacian-based smoothness prior for soft decoding of JPEG images," *IEEE Transactions on Image Processing*, no. 2, pp.509-524, 2016.

[10] Liu, Jing, Guangtao Zhai, Anan Liu, Xiaokang Yang, Xibin Zhao, and Chang Wen Chen, "IPAD: Intensity potential for adaptive de-quantization." *IEEE Transactions on Image Processing*, no. 10, pp. 4860-4872, 2018.

[11] Liu, Jing, Wanning Sun, and Yutao Liu, "Bit-depth enhancement via convolutional neural network," In *International Forum on Digital TV and Wireless Multimedia Communications*, Springer, Singapore, 2017, pp. 255-264.

[12] Zhao, Yang, Ronggang Wang, Wei Jia, Wangmeng Zuo, Xiaoping Liu, and Wen Gao. "Deep Reconstruction of Least Significant Bits for Bit-Depth Expansion," *IEEE Transactions on Image Processing*, 2019.

[13] Hou, Xianxu, and Guoping Qiu, "Image companding and inverse halftoning using deep convolutional neural networks," *arXiv preprint arXiv:1707.00116, 2017*.

[14] Eilertsen, Gabriel, Joel Kronander, Gyorgy Denes, Rafał K. Mantiuk, and Jonas Unger, "HDR image reconstruction from a single exposure using deep CNNs," *ACM Transactions on Graphics (TOG)*, no. 6, pp.178, 2017.

[15] Hao, Hu, Yang Zhang, Dimitris Agrafiotis, Matteo Naccari, and Marta Mrak, "Performance evaluation of reverse tone mapping operators for dynamic range expansion of SDR video content," in *IEEE 19th International Workshop on Multimedia Signal Processing (MMSP)*, IEEE, 2017, pp. 1-6.

[16] Goodfellow, Ian, Jean Pouget-Abadie, Mehdi Mirza, Bing Xu, David Warde-Farley, Sherjil Ozair, Aaron Courville, and Yoshua Bengio, "Generative adversarial nets." In *Advances in neural information processing systems*, , 2014, pp. 2672-2680.

[17] Zhao, Lijun, Huihui Bai, Jie Liang, Bing Zeng, Anhong Wang, and Yao Zhao, "Simultaneous color-depth super-resolution with conditional generative adversarial networks," *Pattern Recognition*, pp. 356-369, 2019.

[18] Zhu, Jun-Yan, Taesung Park, Phillip Isola, and Alexei A. Efros, "Unpaired image-to-image translation using cycle-consistent adversarial networks," In *IEEE international conference on computer vision*, 2017, pp. 2223-2232.

[19] Taigman, Yaniv, Adam Polyak, and Lior Wolf, "Unsupervised cross-domain image generation," *arXiv preprint arXiv:1611.02200*, 2016,.

[20] Liu, Ming-Yu, and Oncel Tuzel, "Coupled generative adversarial networks," In *Advances in neural information processing systems*, 2016, pp. 469-477.

[21] Brock, Andrew, Jeff Donahue, and Karen Simonyan, "Large scale gan training for high fidelity natural image synthesis," *arXiv preprint arXiv:1809.11096*, 2018.

[22] Wang, Chaoyue, Chang Xu, Chaohui Wang, and Dacheng Tao, "Perceptual adversarial networks for image-to-image transformation," *IEEE Transactions on Image Processing*, no. 8, pp. 4066-4079, 2018.

[23] Ledig, Christian, Lucas Theis, Ferenc Huszár, Jose Caballero, Andrew Cunningham, Alejandro Acosta, Andrew Aitken et al, "Photo-realistic single image super-resolution using a generative adversarial network," In *IEEE conference on computer vision and pattern recognition*, 2017, pp. 4681-4690.

[24] Sønderby, Casper Kaae, Jose Caballero, Lucas Theis, Wenzhe Shi, and Ferenc Huszár, "Amortised map inference for image super-resolution," *arXiv preprint arXiv:1610.04490*, 2016.

[25] Reed, Scott E., Zeynep Akata, Santosh Mohan, Samuel Tenka, Bernt Schiele, and Honglak Lee. "Learning what and where to draw." In *Advances in Neural Information Processing Systems*, 2016, pp. 217-225.

[26] Reed, Scott, Zeynep Akata, Xinchen Yan, Lajanugen Logeswaran, Bernt Schiele, and Honglak Lee, "Generative adversarial text to image synthesis," *arXiv preprint arXiv:1605.05396*, 2016.

[27] Zhang, Han, Tao Xu, Hongsheng Li, Shaoting Zhang, Xiaogang Wang, Xiaolei Huang, and Dimitris N. Metaxas, "Stackgan: Text to photo-realistic image synthesis with stacked generative adversarial networks," In *IEEE International Conference on Computer Vision*, 2017, pp. 5907-5915.

[28] Shen, Wei, and Rujie Liu, "Learning residual images for face attribute manipulation," In *IEEE Conference on Computer Vision and Pattern Recognition*, 2017, pp. 4030-4038.

[29] Ranjan, Rajeev, Vishal M. Patel, and Rama Chellappa, "Hyperface: A deep multi-task learning framework for face detection, landmark localization, pose estimation, and gender recognition," *IEEE Transactions on Pattern Analysis and Machine Intelligence*, no. 1, pp. 121-135, 2019.

[30] Lai, Wei-Sheng, Jia-Bin Huang, Narendra Ahuja, and Ming-Hsuan Yang, "Fast and accurate image super-resolution with deep laplacian pyramid networks," *IEEE transactions on pattern analysis and machine intelligence*, 2018.

[31] Iizuka, Satoshi, Edgar Simo-Serra, and Hiroshi Ishikawa, "Globally and locally consistent image completion," *ACM Transactions on Graphics*, no. 4, pp, 107, 2017.

[32] Radford, Alec, Luke Metz, and Soumith Chintala, "Unsupervised representation learning with deep convolutional generative adversarial networks," *arXiv preprint arXiv:1511.06434* , 2015.

[33] Karras, Tero, Timo Aila, Samuli Laine, and Jaakko Lehtinen, "Progressive growing of gans for improved quality, stability, and variation," *arXiv preprint arXiv:1710.10196*, 2017.

[34] Arjovsky, Martin, Soumith Chintala, and Léon Bottou, "Wasserstein generative adversarial networks," In *International Conference on Machine Learning*, 2017, pp. 214-223.

[35] Borji and Ali, "Pros and cons of gan evaluation measures," *Computer Vision and Image Understanding*, pp. 41-65, 2019.

[36] Lei, Na, Kehua Su, Li Cui, Shing-Tung Yau, and Xianfeng David Gu, "A geometric view of optimal transportation and generative model," *Computer Aided Geometric Design*, pp: 1-21, 2019.

[37] Arbel, Michael, Dougal Sutherland, Mikołaj Bińkowski, and Arthur Gretton, "On gradient regularizers for MMD GANs," In *Advances in Neural Information Processing Systems*, 2018, pp. 6700-6710.

[38] Gulrajani, Ishaan, Faruk Ahmed, Martin Arjovsky, Vincent Dumoulin, and Aaron C. Courville, "Improved training of wasserstein gans," In *Advances in Neural Information Processing Systems*, 2017, pp. 5767-5777.

[39] Odena, Augustus, Christopher Olah, and Jonathon Shlens, "Conditional image synthesis with auxiliary classifier gans," In *34th International Conference on Machine Learning*, 2017, pp. 2642-2651.

[40] Lai, Danyu, Wei Tian, and Long Chen. "Improving classification with semi-supervised and fine-grained learning." *Pattern Recognition*, pp. 547-556, 2019.

[41] Bahdanau, Dzmitry, Kyunghyun Cho, and Yoshua Bengio, "Neural machine translation by jointly learning to align and translate," *arXiv preprint arXiv:1409.0473*, 2014.

[43] Gregor, Karol, Ivo Danihelka, Alex Graves, Danilo Jimenez Rezende, and Daan Wierstra, "Draw: A recurrent neural network for image generation," *arXiv preprint arXiv:1502.04623*, 2015.

[43] Xu, Kelvin, Jimmy Ba, Ryan Kiros, Kyunghyun Cho, Aaron Courville, Ruslan Salakhudinov, Rich Zemel, and Yoshua Bengio, "Show, attend and tell: Neural image caption generation with visual attention," In *International conference on machine learning*, 2015, pp. 2048-2057.

[44] Yang, Zichao, Xiaodong He, Jianfeng Gao, Li Deng, and Alex Smola, "Stacked attention networks for image question answering," In *IEEE conference on computer vision and pattern recognition*, 2016, pp. 21-29.

[45] Parikh, Ankur P., Oscar Täckström, Dipanjan Das, and Jakob Uszkoreit, "A decomposable attention model for natural language inference," *arXiv preprint arXiv:1606.01933*, 2016.

[46] Cheng, Jianpeng, Li Dong, and Mirella Lapata, "Long short-term memory-networks for machine reading," *arXiv preprint arXiv:1601.06733*, 2016.

[47] Peng, Yuxin, and Jinwei Qi, "Cm-gans: Cross-modal generative adversarial networks for common representation learning," *ACM Transactions on Multimedia Computing, Communications, and Applications*, no. 1, pp. 22, 2019.

[48] Gu, Jiuxiang, Zhenhua Wang, Jason Kuen, Lianyang Ma, Amir Shahroudy, Bing Shuai, Ting Liu et al, "Recent advances in convolutional neural networks," *Pattern Recognition*, pp. 354-377, 2018.

[49] Wang, Xiaolong, Ross Girshick, Abhinav Gupta, and Kaiming He. "Non-local neural networks." In Proceedings of the IEEE Conference on Computer Vision and Pattern Recognition, pp. 7794-7803. 2018.

[50] Wu, Yan, Yajun Ma, Jing Liu, Jiang Du, and Lei Xing, "Self-attention convolutional neural network for improved MR image reconstruction," *Information Sciences*, pp. 317-328, 2019.

[51] Vaswani, Ashish, Noam Shazeer, Niki Parmar, Jakob Uszkoreit, Llion Jones, Aidan N. Gomez, Łukasz Kaiser, and Illia Polosukhin, "Attention

is all you need," In *Advances in neural information processing systems*, 2017, pp. 5998-6008.

[52] Zhang, Han, Ian Goodfellow, Dimitris Metaxas, and Augustus Odena, "Self-attention generative adversarial networks," *arXiv preprint arXiv:1805.08318*, 2018.

[53] Qian, Rui, Robby T. Tan, Wenhan Yang, Jiajun Su, and Jiaying Liu, "Attentive generative adversarial network for raindrop removal from a single image," In *IEEE Conference on Computer Vision and Pattern Recognition*, 2018, pp. 2482-2491.

[54] Chen, Chen, Qifeng Chen, Jia Xu, and Vladlen Koltun, "Learning to see in the dark," In *IEEE Conference on Computer Vision and Pattern Recognition*, 2018, pp. 3291-3300.

[55] Lim, Bee, Sanghyun Son, Heewon Kim, Seungjun Nah, and Kyoung Mu Lee, "Enhanced deep residual networks for single image super-resolution," In *IEEE Conference on Computer Vision and Pattern Recognition Workshops*, 2017, pp. 136-144.

[56] Yu, Xin, and Fatih Porikli, "Imagining the unimaginable faces by deconvolutional networks," *IEEE Transactions on Image Processing*, no. 6, pp. 2747-2761, 2018.

[57] Yu, Xin, and Fatih Porikli, "Hallucinating very low-resolution unaligned and noisy face images by transformative discriminative autoencoders," In *IEEE Conference on Computer Vision and Pattern Recognition*, 2017, pp. 3760-3768.

[58] Xu, Xiangyu, Deqing Sun, Jinshan Pan, Yujin Zhang, Hanspeter Pfister, and Ming-Hsuan Yang, "Learning to super-resolve blurry face and text images," In *IEEE International Conference on Computer Vision*, 2017, pp. 251-260.

[59] Simonyan, Karen, and Andrew Zisserman, "Very deep convolutional networks for large-scale image recognition," *arXiv preprint arXiv:1409.1556*, 2014.

[60] Zhang, Zhendong, Xinran Wang, and Cheolkon Jung, "DCSR: Dilated Convolutions for Single Image Super-Resolution," *IEEE Transactions on Image Processing*, no. 4, pp.1625-1635, 2019.

[61] Johnson, Justin, Alexandre Alahi, and Li Fei-Fei, "Perceptual losses for real-time style transfer and super-resolution," In *European conference on computer vision*, Springer, Cham, 2016, pp. 694-711.

[62] Everingham, Mark, and John Winn, "The PASCAL visual object classes challenge 2012 (VOC2012) development kit," *Pattern Analysis, Statistical Modelling and Computational Learning*, 2011.

[63] Dong, Chao, Chen Change Loy, Kaiming He, and Xiaoou Tang, "Image super-resolution using deep convolutional networks," *IEEE transactions on pattern analysis and machine intelligence*, no. 2, pp. 295-307, 2015.

[64] Weber, Allan G, "The USC-SIPI image database version 5," *USC-SIPI Report*, 1997, pp. 1-24.

[65] Xin Yu, Basura Fernando, Bernard Ghanem, Fatih Porikli, and Richard Hartley. Face super-resolution guided by facial component heatmaps. In ECCV, pages 217–233, 2018.

[66] Xin Yu, Basura Fernando, Richard Hartley, and Fatih Porikli. Super-resolving very low-resolution face images with supplementary attributes. In Proceedings of the IEEE Conference on Computer Vision and Pattern Recognition (CVPR), pages 908–917, 2018.

[67] Xin Yu and Fatih Porikli. Ultra-resolving face images by discriminative generative networks. In ECCV, pages 318–333, 2016.

[68] Xin Yu and Fatih Porikli. Face hallucination with tiny unaligned images by transformative discriminative neural networks. In AAAI, 2017.

[69] Xin Yu, Fatih Porikli, Basura Fernando, and Richard Hartley. Hallucinating unaligned face images by multiscale transformative discriminative networks. International Journal of Computer Vision, 128(2):500–526, 2020.

[70] Xin Yu, Fatemeh Shiri, Bernard Ghanem, and Fatih Porikli. Can we see more? joint frontalization and hallucination of unaligned tiny faces. IEEE transactions on pattern analysis and machine intelligence, 2019.

[71] Xin Yu, Basura Fernando, Richard Hartley, and Fatih Porikli. Semantic face hallucination: Super-resolving very low-resolution face images with supplementary attributes. IEEE Transactions on Pattern Analysis and Machine Intelligence, 2019.